\documentclass[sigconf]{acmart}

\usepackage{tabularx}
\usepackage{stfloats}
\usepackage{graphicx}
\usepackage{caption}

\AtBeginDocument{%
  \providecommand\BibTeX{{%
    \normalfont B\kern-0.5em{\scshape i\kern-0.25em b}\kern-0.8em\TeX}}}

\setcopyright{acmcopyright}
\copyrightyear{2024}
\acmYear{2024}
\acmDOI{10.1145/1122445.1122456}

\acmJournal{JACM}
\acmVolume{37}
\acmNumber{4}
\acmArticle{111}
\acmMonth{8}

\graphicspath{{figures/}}

\begin{document}

\title[Cosplay Commission]{Collective Creation of Intimacy: Exploring the Cosplay Commission Practice within the Otome Game Community in China}

\author{Yihao Zhou}
\orcid{0000-0002-2947-6243}
\affiliation{%
  \institution{Pennsylvania State University}
  \city{University Park}
  \state{Pennsylvania}
  \country{USA}
}
\email{yihaozhou@psu.edu}

\author{Haowei Xu}
\orcid{0009-0002-6014-755X}
\affiliation{%
  \institution{Open Lab, Newcastle University}
  \city{Newcastle upon Tyne}
  \country{United Kingdom}
}
\email{haowei.xu@newcastle.ac.uk}

\author{Lili Zhang}
\orcid{0009-0007-7214-8943}
\affiliation{%
  \institution{Temple University}
  \city{Philadelphia}
  \state{Pennsylvania}
  \country{USA}
}
\email{lili.zhang@temple.edu}

\author{Shengdong Zhao}
\orcid{0000-0001-7971-3107}
\affiliation{%
  \institution{City University of Hong Kong}
  \city{Hong Kong}
  \country{China}
}
\email{shengdong.zhao@cityu.edu.hk}


\renewcommand{\shortauthors}{Zhou et al.}

\begin{abstract}
Cosplay commission (cos-commission) is a new form of commodified intimate relationship within the Otome game community in China. To explore the motivations, practices, experiences, and challenges, we conducted semi-structured interviews with 15 participants in different roles. Our findings reveal that cos-commission, as a hybrid activity, provides participants with a chance to collaboratively build meaningful connections. It also offers a pathway for personal exploration and emotional recovery. However, the vague boundary between performative roles and intimate interactions can give rise to unexpected negative outcomes, such as attachment-driven entanglements and post-commission ``withdrawal symptoms.'' While digital platforms facilitate communication in cos-commissions, they often lack sufficient safeguards. This preliminary work provides insights into the formation process of hybrid intimate relationship and its potential to foster personalized, long-term support for mental well-being, and reveals potential privacy and safety challenges.
\end{abstract}


\begin{teaserfigure}
\centering
  \includegraphics[width=0.9\textwidth]{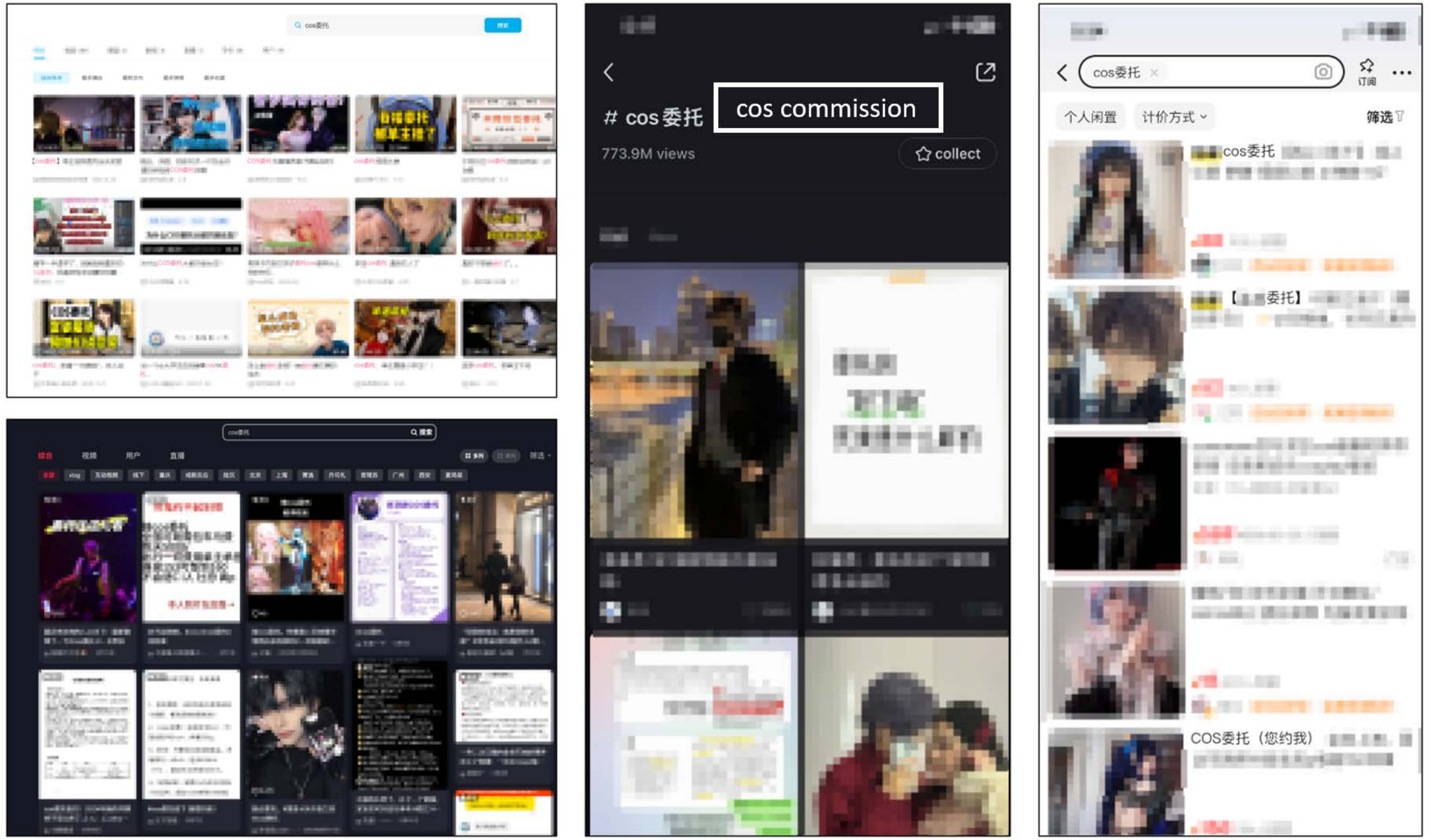}
  \caption{Cosplay commission-related posts on different digital platforms (BiliBili, Douyin, Xiaohongshu and
Xianyu).}
  \Description{Posts of cosplay commissions on different digital platforms.} 
  \label{fig:teaser}
\end{teaserfigure}
\maketitle

\balance

\section{Introduction}
In recent years, Otome games (story-based romance video games for female players) have formed a large presence in the subcultural community of China. Players spend their time and money to gain access to storylines and to form relationships with virtual characters. Recently, the offline community is even growing further, with more and more players and game developers hosting meetups in different areas. In tandem with this trend, a new form of commodified intimate experience has emerged, which referred to players as ``cosplay commission'' or cos-commission in short. Through cos-commission, players can hire cosplayers to physically represent the game characters and have a "one-day romance" experience, thus bringing their virtual interactions into reality as they can .

The rise of cos-commission has received attention from both academia~\cite{lei_2024_otome, Ganzon_2019b, wu_2024_romantic} and the media~\cite{wu_2023_china, zhu_2024_china, feng_2018, liang_2019}, reflecting the rapid speed of its growth. For HCI researchers, online game communities have long been one of the main areas. Previous works on Massively Multiplayer Online Games (MMOGs) have demonstrated how virtual worlds can create shared social norms among a group of people~\cite{boellstorff2015coming, nardi2010my, zhong_2011_collective, Kou2014playing, nick_2006_motivation} along with fostering meaningful social relationships and social support~\cite{tong_2021_animal, freeman_2016_revisiting, lei_2024_otome}. These findings lay a strong theoretical foundation for the development and study of how game communities as a whole contribute to collective in-game goals. However, empirical knowledge on how social cooperation is organized at the level of individual interactions in order to address players' unique personal needs is limited. In contrast to MMOGs that emphasize sociality and cooperation~\cite{nick_2006_motivation}, the experiences of Otome players are more personal and customized. Accordingly, cos-commission, as a player-motivated small practice, provides a useful lens through which to investigate how players work together to make sense of the game in a more individual game environment. 

Based on semi-structured interviews with 15 players within the cos-commission ecosystem, we find that this practice allows players to co-create personal intimate experiences and to recover from negative emotions. Cos-commissions offer players freedom through crafted encounters. However, risks such as post-commission "withdrawal symptoms" and blurred barriers between performative roles and actual emotions demonstrate a need for safeguards and supports. Though digital platforms and tools enable hybrid interaction in cos-commission, the current designs are in short of approaches to handle the risks. As such, we argue for features on digital platforms that enable creative expression and exposure control, thus allowing subculture communities to grow safely.

\section{Background and Related Work}
Otome game, sometimes contracted to Otoge, is a story-based video game genre that is targeted towards female players~\cite{tian_2022_otome}. The first commercial Otome game, \textit{Angelique}, appeared in the early 1990s in Japan~\cite{Kim2009-KIMWGI}. As China's mobile gaming market exploded during the 2010s, the Chinese market of Otome games expanded by leaps and bounds and gained a huge number of fans.veral mobile Otome games that achieved notable commercial success were launched in this period, such as \textit{Mr. Love: Queen's Choice}\footnote{https://evol.papegames.cn/ [Last Accessed: Nov 30, 2024]}, \textit{Light and Night}\footnote{https://love.qq.com/ [Last Accessed: Nov 30, 2024]}, \textit{Tears of Themis}\footnote{https://tot.hoyoverse.com/ [Last Accessed: Nov 30, 2024]}, and \textit{Love and Deepspace}\footnote{https://loveanddeepspace.infoldgames.com/ [Last Accessed: Nov 30, 2024]}. Alongside the flourish of market, offline events including fan meet-ups and dedicated conventions, have become common.

From a sociology perspective, subcultures can be described as social groups that diverge from the mainstream through unique practices, signs, and shared identities~\cite{haenfler2013subcultures}. Contemporary sociology scholars proposed the post-subcultural theory \cite{williams2011subcultural}, which explained that subcultures are more fragmented and more fluid and hybrid in the present, with various practices and participants. In HCI, subcultures have been examined for their distinct communities in relation to technology use, social norms for participation, and technology generation~\cite{Tanenbaum2018SteampunkSA, su_2019_dolls, su2019speedrun}. Cosplay is a good example of this hybridization, where individual identity and collective identity are fused together through the performance of using inputs from anime, comics, games, and novels (ACGN)~\cite{rahmen_2012_cosplay, ito_2012_fandom}.

Within the context of the purpose of cos-commission, ``intimacy'' is a complex goal and consists of emotional affinity, mutual comprehension, and shared experience. Members of the Otome game communities develop intimacy in different mediated ways online – from exchanging photos to joint participation in tasks or shared discussions. The advantage of online intimacy over offline intimacy is multi-folded. There is evidence that online intimacy is associated with higher levels of social support for minority members who may experience greater difficulties in creating close relationships offline~\cite{jeffrey2014raccoon,freeman_2016_revisiting}. In terms of accessibility, digital systems have been noticed for their unique affordances for cultivating intimacy, for example, anonymity, decentralization, and shared gradual goals~\cite{freeman_2016_revisiting, li_2023_virtual, maloney_falling_2020, boellstorff2015coming}. Yet little is known about how the co-creation of intimate experiences takes place in hybrid settings that online with offline experiences. Cos-commission as a node between online subcultural community practices and the co-creation of intimacy offline, offers a useful framing to answer this question.

\section{Methodology}
\begin{table*}[htbp]
    \centering
    \caption{Demographic Information of Interviewees}
    \label{demographic}
    \begin{tabularx}{0.7\textwidth}{c|c|c|c|c|c}
        \toprule
        \textbf{ID} & \textbf{Age} & \textbf{Gender} & \textbf{Education} & \textbf{Role in cos-commission} & \textbf{Experience}\\
        \midrule
        P1  & 21 & Female  & Bachelor's & Cosplayer & 12 months\\ 
        P2  & 22 & Female  & Bachelor's & Customer & 24 months\\ 
        P3  & 18 & Female  & High School & Customer & 10 months\\
        P4  & 22 & Female  & Bachelor's & Cosplayer & 12 months\\ 
        P5  & 20 & Female  & Bachelor's & Both & 6 months\\ 
        P6  & 21 & Male    & Bachelor's & Cosplayer & 12 months\\
        P7  & 19 & Female  & Bachelor's & Both & 13 months\\ 
        P8  & 20 & Female  & Bachelor's & Cosplayer & 24 months\\ 
        P9  & 19 & Female  & Bachelor's & Customer & 13 months\\
        P10  & 21 & Female  & Bachelor's & Both & 14 months\\ 
        P11  & Anonymous & Anonymous  & Anonymous & Anonymous & Anonymous\\ 
        P12  & 21 & Female  & Bachelor's & Customer & 10 months\\
        P13  & 20 & Female  & Bachelor's & Both & 13 months\\ 
        P14  & 22 & Female  & Bachelor's & Both & 12 months\\ 
        P15  & 22 & Female  & Bachelor's & Both & 15 months\\
        \bottomrule
    \end{tabularx}
\end{table*}

To understand the motivations, interactions, and challenges experienced by both customers and cosplayers in the cos-commission ecosystem, as well as their expectations for future practice, we conducted semi-structured interviews with 15 participants in China who had participated in cos-commission either as cosplayers or customers (14 females and 1 male). The demographic information of our participants is detailed in Table \ref{demographic}. Participants were recruited via snowball sampling. Prior to the interviews, they were briefed on the study's purpose and background. With their consent, all interviews were recorded and transcribed verbatim in Chinese. Interview questions covered participants' past experiences with cos-commission, their motivations, preparations, and reflections on the experience. Follow-up questions included challenges encountered in cos-commission, potential improvements, and their opinions on the future of practice. Interviews averaged about 60 minutes, and each participant was compensated 60 CNY. Two native Mandarin-speaking authors independently coded 20\% of the transcripts and then met to establish a unified codebook. The codes in the codebook were than iteratively organized into a thematic hierarchy~\cite{clarke2017thematic} and resulted in a consensus on the motivations, practices, challenges, and safety/privacy concerns of participants, as detailed in the following section.

\section{Findings}
In this section, we will report our findings from the interviews, including the role-specific motivations, interaction dynamics, and risks along with cos-commissions.

\subsection{Role-Specific Motivations}

Our preliminary investigation reveals that motivations vary based on participants' roles in the cos-commission ecosystem and their past experiences. There are also some shared motivations, such as social and emotional needs. However, cosplayers have different expectations compared with customers. Specifically, cosplayers pay more attention to curating identity through performative labor, while customers focus more on fulfilling their social and emotional needs.

\subsubsection{Cosplayers}

Although some cosplayers describe cos-commission as a part-time job, most do not expect considerable financial rewards from this activity. Many expressed ethical considerations in pricing, recognizing that most customers are students with limited income. For instance, P1 mentioned, \textit{``Typically, these young girls (customers) don't have much money, so I won't set my prices too high.''} For these cosplayers, the price is perceived as compensation for their time and preparation of character costumes and make-up rather than a salary.

Besides financial reward, many cosplayers are driven by curiosity and a desire to pursue personal interests when accepting commissions. For example, P7, an experienced cosplayer, shared with us, \textit{``My first encounter with cos-commission was at a Comic Convention, where someone saw me cosplaying a character she liked and asked if I took commission requests. Since I am also an Otome game player, I wanted to try it out. That's how I started accepting commissions on different platforms.''} This interest-driven motivation is further reflected in cosplayers' preference for commissioning characters they are familiar with. As P7 described, \textit{``If you go on a date with someone unfamiliar with the character, you'll get a normal dating experience, but it won't evoke the same depth of feeling as it would with someone who truly understands the character.''} For these cosplayers, financial reward is seen as an additional benefit—i.e., a reward for their skill and passion—rather than the primary goal.

\subsubsection{Customers}

Motivations for customers can be quite complex as their backgrounds and personal goals vary. Overall, the commonly shared motivators are divided into two categories: social motivators and emotional motivators.

\paragraph{Social Motivator}
The social motivator consumers have for cos-commission is an example of the interplay between cos-commission activities and social dynamic at the level of consumer practices, popular culture, and social life. For example, P2 highlighted a coincidence of seeing cos-commissioning parties when attending conventions: \textit{``Attending conventions with friends has been a primary motivation for me. As Otome games grew in popularity, so did conventions and cos-commission. At a convention, I saw a single customer hire six cosplayers simultaneously for group photos. That is when I first became aware of it.''} Offline events present these customers with a wealth of opportunities to grab onto Otome game cultures and at the same time support them in keeping up relationships with their friends and other folk fans. Likewise, P3 described her first exposure to cos-commission as serendipity when she encountered videos on social media, \textit{``I found other people's cos-commission videos and their sweet interaction. That's when I figured I would give it a shot myself. I wanted to make some memories.''} The increase of cos-commission coincides with the rise of Otome games and related offline events, indicating further growth within the niche space of this subcultural community and approachability to the aspirations of community members.

\paragraph{Emotional Motivator}
Many customers view cos-commission as a continuation of the in-game emotional satisfaction. For example, P2 told us, \textit{``He (the virtual character) is so perfect. Over time, I developed a desire to touch him, even though it was impossible. I want to hug him, hold his hands, and wish I could meet him in real life.''} Through cos-commission, they form an opportunity to have physiological or closer communications that virtual worlds can not offer.

Others struggling with mental health find cos-commission a source of emotional recovery and support otherwise missing from their lives. For example, P9 described cos-commission as the road to recovery: \textit{``I wish I could get someone who resembles my lover to play him in real life. It would be a reward for me."} Another participant, P2, in contrast, took cos-commission to be an act of introspection: \textit{"I want to verify through cos-commission that I have the ability to love and communicate.''} Despite that there is no concrete interaction rule, the positive impact of cos-commission to mental health should not be ignored. Many customers who have extensive experience in Otome games or cos-commissions reported having their own favorite character. Cos-commission allows them to have physical contact with their "dream lover" as part of an attempt to turn digital connections into real ones.

\subsection{Progressive Intimate Experience}

In this subsection, we discuss a collective intimacy-building process that evolved in the cos-commission practice.

\subsubsection{Pre-commission}

Selective matching and boundary negotiation are commonly seen as a first step in the pre-commission phase. Before the commission took place, customers filter cosplayers with several criteria. Although the details are different, there are some shared concerns. To streamline the process and make the criteria clear, many customers and cosplayers have prepared a short questionnaire, including questions about physical appearance/BMI stats, etc., as well as queries on nickname, level of intimacy that one is comfortable with, and relationship status. Besides these standard points, there are some more vague criteria. For example, P14 considered the impression she had of the character to identify which cosplayers are more likely to be chosen: \textit{``The appearance of the cosplayer needs to be consistent with the impression I have of the character.''} Among our participants, we find that physical appearance is one of the most cited reasons when selecting a cosplayer.

\subsubsection{During the Cos-commission}

During a typical cos-commission, there will be a progressive relationship development with planned interactions. Dating places and activities are usually decided by the customer. In the case that the customer cannot or do not want to make that decision, it is the cosplayer who will do so instead. Several customers told us it feels like the dating gets better as the day goes on, that they start a little bit awkward but have a close connection by the end. P3 talked about this feeling in terms of nicknames: \textit{``In the beginning, I called her 'teacher' where we didn't know each other, and she called me 'sponsor'. As we got closer, I used more endearing, couple-style names or cute nicknames, such as `little xxx'.''} This change reflects an intimacy evolution from established comfort and trust.

\subsubsection{Post-commission}
\label{post-commission}
After commission, participants often have some rituals to document the experience and keep emotional attachments at bay. For some, they will prepare detailed feedback and gifts, including handmade thank-you letters or mementos of gratitude. Respectively, we have observed a long-term emotional fulfillment from these exchanges. According to P4, \textit{``After the commission, the cosplayer said encouraging words to me. If someone else had said it, I might not have listened, but because it was from my cosplayer, I did listen. ... We promised to meet again one day. When I see the photos or other mementos from that day, I remember the promise, which gives me strength to keep going.''} This long-term impact is counter-intuitive to the temporary nature of cos-commission and provides evidence for the power of offline exchanges expanding the benefit of a virtual world.

\subsection{Privacy and Safety Risks}
While cos-commission offers positive experiences, there are some underlying risks. In this subsection, we describe some common issues.

\subsubsection{Withdrawal Symptoms}
From our interviews, we noticed that almost all customers reported experiencing varying degrees of what they called ``\textit{withdrawal symptoms}'' afterward. This term is used to describe states of emotional confusion and difficulty readjusting to everyday life at the end of the commission. P3 likened cos-commission to a dream: \textit{``There is a feeling that what I experienced that day was like a dream, much like Cinderella's crystal shoes gave you a beautiful dream. When the day ends, you have to go back. It's very painful.''} While some post-commission rituals have been adopted to minimize participants' emotional attachment, as discussed in section \ref{post-commission}, the methods still may not be sufficient to allow participants to fully swipe out the negative emotional impact of cos-commissions. As a result, long-term emotional support is weakened by the feeling of loss and contrast. 

Immersive environments have the ability to evoke deeper emotions and empathy~\cite{liu2022high,liu2024attention,you_2023_bulevr}. However, in cos-commission, unlike formal psychotherapy~\cite{landy_1994_drama}, there is a lack of an exit mechanism, and the participants are not well-trained. Consequently, the deep, extensive, and dyadic interactions within a short period cause difficulties for participants in digesting and processing their emotions.

\subsubsection{Blurred Boundaries}
Currently, there is no structured guideline for cos-commission. Cosplayers typically self-learn the personality of characters by watching anime, manga, and videos to provide an immersive and high-quality service. However, unlike other forms of paid companionship, such as sex workers~\cite{Saha_2024_sex}, paid game teammates~\cite{shen_2021_labor}, virtual lovers~\cite{li_2023_virtual} or host/hostess clubs in Japan~\cite{takeyama2020staged}, cos-commission participants do not share a clear, unified understanding of their short-term relationship, which may lead to unexpected and unavoidable emotional transference. As such, it often becomes difficult for participants to distinguish whether their feelings are towards the virtual character, the cosplayer, or their imaginations, leaving a paradoxical situation where the performance ends but the emotional immersion does not.

\subsubsection{Unwanted Contact and Information Leakage}
Digital platforms, as illustrated in Figure \ref{fig:teaser}, are the main venues for organizing cos-commissions. However, they also present some vulnerabilities.

On different platforms, participants have a variety of options to customize their media preferences. Video-sharing websites, such as BiliBili~\footnote{BiliBili: A Chinese online video-sharing platform. \url{https://bilibili.com/} [Last Accessed: May 17, 2025]}, provide open spaces for users to showcase positive experiences. Xiaohongshu~\footnote{Xiaohongshu: Known in English as RedNote, a Chinese social networking and e-commerce platform. \url{https://xiaohongshu.com/} [Last Accessed: May 17, 2025]} (RedNote), with a fusion of graphic and textual information, enables customers to search for cosplayers, comment on reviews and service authenticity. Douyin~\footnote{Douyin: A short-video platform in mainland China and Hong Kong. While TikTok and Douyin share a similar interface, they operate separately. \url{https://douyin.com/} [Last Accessed: May 17, 2025]}, which is called TikTok outside mainland China and Hong Kong, specializes in short videos, allows customers to verify the skills of cosplayers via their previous performances. Xianyu~\footnote{Xianyu: An online second-hand trading platform. \url{https://goofish.com/} [Last Accessed: May 17, 2025]}, a second-hand trading platform, simplifies the cos-commission process with quick communication and payments and has many who are convenience-minded users. By jumping back-and-forth between different platforms, customers can find their preferred cosplayers and improve their cos-commission experiences with easy, safe, and high-quality transactions.

Despite the benefits, there are safety and privacy concerns over digital platforms. Professional cosplayers express worries about being harassed or subjected to entanglements from a customer with whom they have built a relationship after a commission ends. For example, P10 reported that someone he knows achieved this: \textit{``My friend posted a photo with her location on her birthday, and a customer followed the location to find her. She didn't think the customer would, so she didn't worry. ... The customer arrived at her birthday party uninvited and insisted she drive him/her home, causing trouble for her and her girlfriend.''} This case reflects a lack of anti-harassment mechanism and profile-management policy in cross-platform settings. Due to the feature of in-depth, dyadic communication and information exchange of cos-commission, cosplayers, and customers are pushed to a higher risk in personally identifiable information disclosure.

\subsubsection{Gender-Related Vulnerability}
Speaking of gender, female cosplayers feared uncomfortable or mistaken encounters with male customers. Some commented that males are in general more physically powerful than females, and therefore if an accident were to happen they would have a harder time defending themselves. In cos-commission, if intimate interactions are to take place, participants sometimes move to private spaces. Since there is no aid to be had from others, their anxiety and vulnerability level will increase even more.

\section{Conclusion and Future work}

In this preliminary study, we explored the motivations, experiences, and issues of the cos-commission participants within the Otome game community in China. Our results shows that cos-commission provides a hybrid space in which customers co-create meaningful connections with cosplayers. It also serves as a medium that allows its participants to rediscover themselves as well as recover from bad emotional states. However, the ambiguous line between the performing roles and real intimacy may result in negative consequences, such as depression caused by attachment and ``withdrawal symptoms'' after performing.

Our investigation also reveals safety and privacy issues associated with the existing cos-commission practice. Vulnerabilities, such as unwanted post-commission entanglements when personally identifiable information (e.g., a phone number, or geolocation data) gets exposed or gender inequality among female cosplayers reflect system-level failures in facilitating cross-platform hybrid intimate interactions. Strong safeguards, such as chat encryption, multi-factor identity verification, and adjustable privacy filters, are required to be put in place by the platforms to help create safer spaces while maintaining the independent creativity of cos-commission culture.

Through our study of subculture communities as a case-in-point, we hope to contribute to the design of virtual world interaction at large. It would be valuable for future research to consider the longitudinal dynamics of cos-commission participation, including the transformation of emotional investments across multiple visits. Together with Otome game communities, we hope to co-design sociotechnical infrastructures that align immersive intimacy with appropriate safeguards. If we (re)valorized emotional satisfaction alongside participant welfare, we might learn from subcultures how to construct a better digital world.



\bibliographystyle{ACM-Reference-Format}
\bibliography{main}


\end{document}